# Experimental Evidence of the Topological Surface States in Mg$_3$Bi$_2$ Films Grown by Molecular Beam Epitaxy

Tong Zhou[1,2,3,4,†], Xie-Gang Zhu[2,4,†], Mingyu Tong[5,†], Yun Zhang[2,4], Xue-Bing Luo[2,4], Xiangnan Xie[1], Wei Feng[2,4], Qiuyun Chen[2,4], Shiyong Tan[2,4], Zhen-Yu Wang[1,3,7,*], Tian Jiang[1,5,*], Xin-Chun Lai[2,*], and Xuejun Yang[1,6]

[1]State Key Laboratory of High Performance Computing, College of Computer, National University of Defense Technology, Changsha 410073, P. R. China
[2]Science and Technology on Surface Physics and Chemistry Laboratory, Jiangyou 621908, Sichuan, China
[3]National Innovation Institute of Defense Technology, Academy of Military Sciences PLA China, Beijing 100010, P. R. China
[4]Institute of Materials, China Academy of Engineering Physics, Mianyang 621700, Sichuan, China
[5]College of Advanced Interdisciplinary Studies, National University of Defense Technology, Changsha 410073, P. R. China
[6]Academy of Military Sciences PLA China, Beijing 100010, P. R. China
[7]Beijing Academy of Quantum Information Sciences, Beijing, 100084, P.R. China

Type-II nodal line semimetal (NLS) is a new quantum state hosting one-dimensional closed loops formed by the crossing of two bands which have the same sign in their slopes along the radial direction of the loop. According to the theoretical prediction, Mg$_3$Bi$_2$ is an ideal candidate for studying the type-II NLS by tuning its spin-orbit coupling (SOC). In this paper, high quality Mg$_3$Bi$_2$ films are grown by molecular beam epitaxy (MBE). By *in-situ* angle resolved photoemission spectroscopy (ARPES), a pair of surface resonance bands (SRBs) around $\bar{\Gamma}$ point is clearly seen. It shows that Mg$_3$Bi$_2$ films grown by MBE is Mg(1)-terminated by comparing the ARPES data with the first principles calculations results. And, the temperature dependent weak anti-localization (WAL) effect in Mg$_3$Bi$_2$ films is observed under low magnetic field, which shows a clear two dimensional (2D) *e-e* scattering characteristics by fitting with the Hikami-Larkin-Nagaoka (HLN) model. Combining with ARPES, magneto-transport measurements and the first principles calculations, this work proves that Mg$_3$Bi$_2$ is a semimetal with topological surface states TSSs, which paves the way for $Mg_3Bi_2$ as an ideal materials platform for studying the exotic features of type-II nodal line semimetals (NLSs) and the topological phase transition by tuning its SOC.

## Introduction

Topological surface states (TSSs) are a class of novel electronic states with great potential for topological quantum computation and spintronics applications. Materials with TSSs, such as topological insulators and topological semimetals have emerged in the past decade as a major research focus in condensed matter physics.[1-10] TSSs are immune to non-magnetic disorders. They eliminate 180° backscattering under the protection of time-reversal symmetry and prevent to be fully gapped when perturbed due to the odd number of band crossing. Besides, due to the time-reversal symmetry, there are two closed electrons scattering paths for surface states. One path accumulates a Berry phase of $\pi$ when across single Dirac cone and induces a destructive quantum interference with the other path, leading to the enhancement of the conductivity. That is the origination of weak antilocalization effect which only exists under low magnetic field for the time-reversal symmetry reason.[11, 12]

NLSs are novel proposed topological solid-state phases which host one-dimensional closed loops or line degeneracies formed by the intersection of two bands. It is proposed as type-II NLSs when the linear spectrum at every point of the nodal line is strongly tilted and tipped over along one transverse direction, which may lead to different magnetic, optical, and transport properties compared with conventional nodal loops. To date, only $K_4P_3$ was theoretically predicted as a type-II NLS, but no candidate material has been experimentally reported. Therefore, it is highly desirable to experimentally realize and study topological materials with clear type-II NLS

signature.[13-15]

Here we investigate $Mg_3Bi_2$, which is predicted as owning TSSs and is an ideal materials platform for studying type-II NLSs.[14,15] Previous studies of $Mg_3Bi_2$ mainly focused on its good thermoelectric properties as one of $Mg_3Sb_2$-class materials.[16,17] From first principles calculations, $Mg_3Bi_2$ was predicted as a three-dimensional (3D) topological insulator with SOC and a type-II NLS without SOC. A topology transition from 3D topological insulator to type-II NLS would be expected by tuning the strength of SOC effect. The previous calculation on $Mg_3Bi_2$ only considered the Mg- and Bi-terminated surface cases, which is insufficient while there are two kinds of Mg atoms distinguished by its structure symmetry.[14,15] Furthermore, ARPES measurements were performed on single crystal $Mg_3Bi_2$ to verify the above theoretical calculations.[15] But due to Mg vacancies, $Mg_3Bi_2$ is a heavy p-type material. Although potassium surface doping had been done, their ARPES results did not show the conduction bands or the surface states critical crossing point. Thus, it is still lacking direct experimental evidence for the TSSs in $Mg_3Bi_2$.

To realize novel topological materials-based devices, such as spin transistors, topological quantum computation, or even to achieve the quantum anomalous Hall effect (QAHE), well-controlled thickness of ultrathin films are inevitable.[18] In this work, we have grown $Mg_3Bi_2$ films by MBE which allows us to control the thickness of $Mg_3Bi_2$ film layer by layer. *In-situ* reflection high energy electron diffraction (RHEED) and *ex-situ* X-ray diffraction (XRD) measurements were performed to confirm the high quality of our $Mg_3Bi_2$ films. First principles calculations and *in-situ* ARPES measurements were performed to study the band structure of $Mg_3Bi_2$. The ARPES spectra are consistent with our first principles calculations, and we confirmed that the natural surface of our films is type-I Mg-terminated, the definition of which will be given below.

To further confirm the topology of its surface states, magneto-transport measurements were performed. We observed clear 2D weak anti-localization (WAL) effect which is the hallmark of topological protected surface states. Therefore, we have provided direct experimental evidences for the existence of TSSs in $Mg_3Bi_2$ film and paved the way to study type-II NLSs.

**Structure characterization of $Mg_3Bi_2$ film**

$Mg_3Bi_2$ crystalizes in a layered Kagome lattice structure with a space group of $P\bar{3}m1$ (No.164). As shown in Figure 1a, the unit cell contains five atomic layers with a stacking sequence of Mg(1)-Bi-Mg(2)-Bi-Mg(1) along the (001) crystallographic orientation, which could be defined as a quintuple layer (QL), similar to that of $A_2B_3$-type topological insulators. Figure 1b depicts the bulk Brillouin zone and the projected surface Brillouin zone (SBZ) of the (001) surface with the high symmetric momentum points marked exclusively. The chemical forces between each two adjacent Mg(1)-Mg(1) layers is van der Waals (vdW) type, which implies the possibilities of synthesizing $Mg_3Bi_2$ films in a layer-by-layer growth mode. We have grown $Mg_3Bi_2$ single crystalline films on graphene which was epitaxially grown on 6H-SiC(0001) substrate. RHEED patterns were collected simultaneously during the growing process, and the sharp streaks along $\bar{\Gamma}-\bar{K}$ and $\bar{\Gamma}-\bar{M}$ directions in the SBZ indicate the high quality of the film (Figure. 1(c) and (d)). XRD clearly revealed the (001), (002), (003), (004) and (005) peaks of $Mg_3Bi_2$ (Figure.1 (f)). The (001) peak intensity is even stronger than that of the SiC(0001) substrate signal, indicating the high crystalline quality of our *as-grown* samples. The in-plane lattice constant deduced from the RHEED streak spacing (Figure 1e) is 4.677Å and the out-of-plane lattice constant calculated from XRD patterns is 7.403Å, which are both consistent with previously reported values.[19]

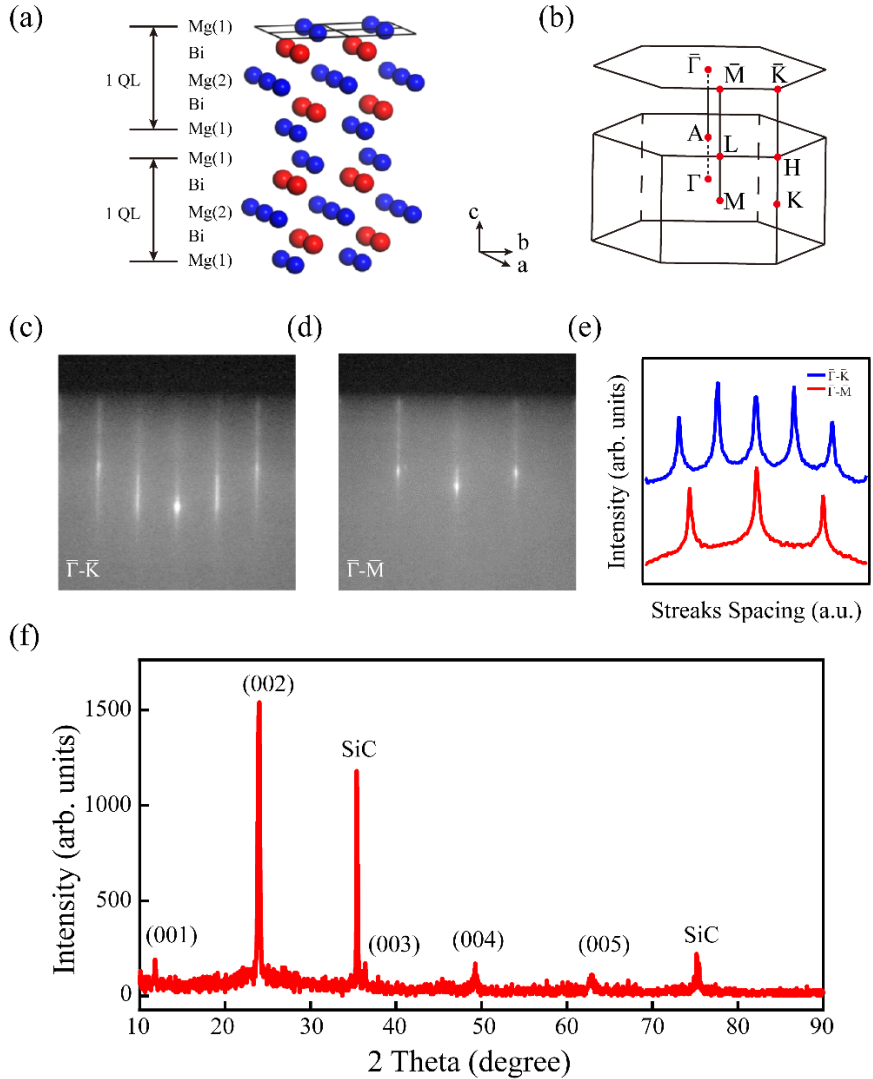

**Figure 1. Characterization of the crystal structure of Mg₃Bi₂ film. (a)** Crystal structure of Mg₃Bi₂. The Mg(1)-Bi-Mg(2)-Bi-Mg(1) quintuple layer (QL) is also schematically shown. **(b)** Bulk Brillouin zone and the projected surface Brillouin zone of the (001) surface. **(c,d)** RHEED patterns of Mg₃Bi₂ film grown on Graphene/6H-SiC(0001) substrate, with the incident electron beam along the $\bar{\Gamma}-\bar{K}$ and $\bar{\Gamma}-\bar{M}$ directions, respectively. **(e)** Spacing of the RHEED streaks in (c) and (d). **(f)** XRD patterns of the Mg₃Bi₂ film, with all the (0,0,1n) diffraction peaks exclusively marked.

**Electronic structure of Mg₃Bi₂ film**

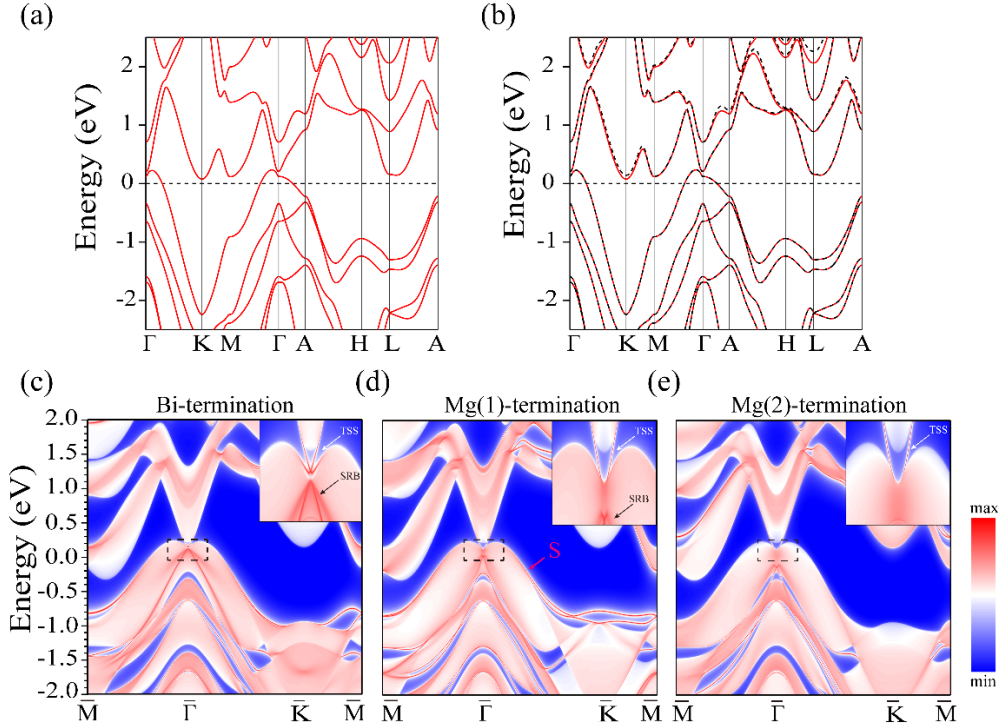

**Figure 2. Theoretical calculations of Mg₃Bi₂.** **(a)** Bulk band structures by first principles calculations. **(b)** Comparison between the bulk band structures by first principles (solid red lines) and tight-binding method (dashed black lines). **(c)-(e)** Band spectra of Bi-, Mg(1)- and Mg(2)-terminated semi-infinite (001) surface slabs. The insets show the detailed electronic structures around the $\bar{\Gamma}$ point near Fermi level.

We have theoretically investigated the electronic structures of Mg₃Bi₂ by adopting the projector augmented wave (PAW) method as integrated in the VASP package. The generalized gradient approximation (GGA) of Perdew-Burke-Ernzerhof functional is used for exchange-correction potential. The Brillouin zone is sampled with 11×11×5 Γ-centered *k*-mesh for the structural optimization and with 15×15×11 *k*-mesh for self-consistent calculations. Spin-orbit coupling effect was fully taken into consideration throughout the whole calculations. The VASP2WANNIER90 interface was used to construct the first principles tight-binding Hamiltonian. Mg $s$ and $p$ and Bi $p$ orbitals are taken into consideration and no maximizing localization procedure was performed to construct the Wannier orbitals.[20-22] The calculations of the surface state spectra and constant energy contours (CECs) were done by making use of the WANNIER-TOOLS packages.[23] Since our grown sample is relatively thick, the surface states and Fermi surface are calculated by tight-binding based Green function method for semi-infinite sample. The bulk band structures are well reproduced by the tight-binding Hamiltonian, as shown in Figure 2a and b. In principle, there are three different possible terminations, i.e., Bi-, Mg(1)- and Mg(2)-terminated surfaces for the (001) surface of $Mg_3Bi_2$, the band spectra of which are demonstrated in Figure 2c,d and e, respectively. From the bulk band spectra, $Mg_3Bi_2$ is a semi-metal with band "overlapping" at $\bar{\Gamma}$ and $\bar{M}$. The non-trivial topological nature of the surface states around $\bar{\Gamma}$ point near the Fermi level has been thoroughly discussed in Ref. 15. Therefore, theoretically, $Mg_3Bi_2$ is a semi-metal with non-trivial topological surface states, similar to the semi-metal Antimony element.[24] The TSS on the Bi-terminated surface possesses a Dirac point (DP) at $\bar{\Gamma}$ inside the local energy gap, while the DPs for the Mg-terminated surfaces are buried in the valence bands. Apart from the TSSs, there are a pair of surface resonance bands (SRBs) around $\bar{\Gamma}$, with different binding energies for the three different terminations. For the Bi-terminated surface, the SRBs share a same peak point at the valence band edge, while there is a Rashba-type splitting between the SRBs for the Mg(1)- and Mg(2)-terminated surfaces. As for the Mg(1)-termination, there are a pair of clear surface states on the brink of the valence band continuum (marked as S), while it is missing for the Mg(2)-termination. The above mentioned features would offer us clear fingerprints for our experimental confirmation of the surface termination in our $Mg_3Bi_2$ films.

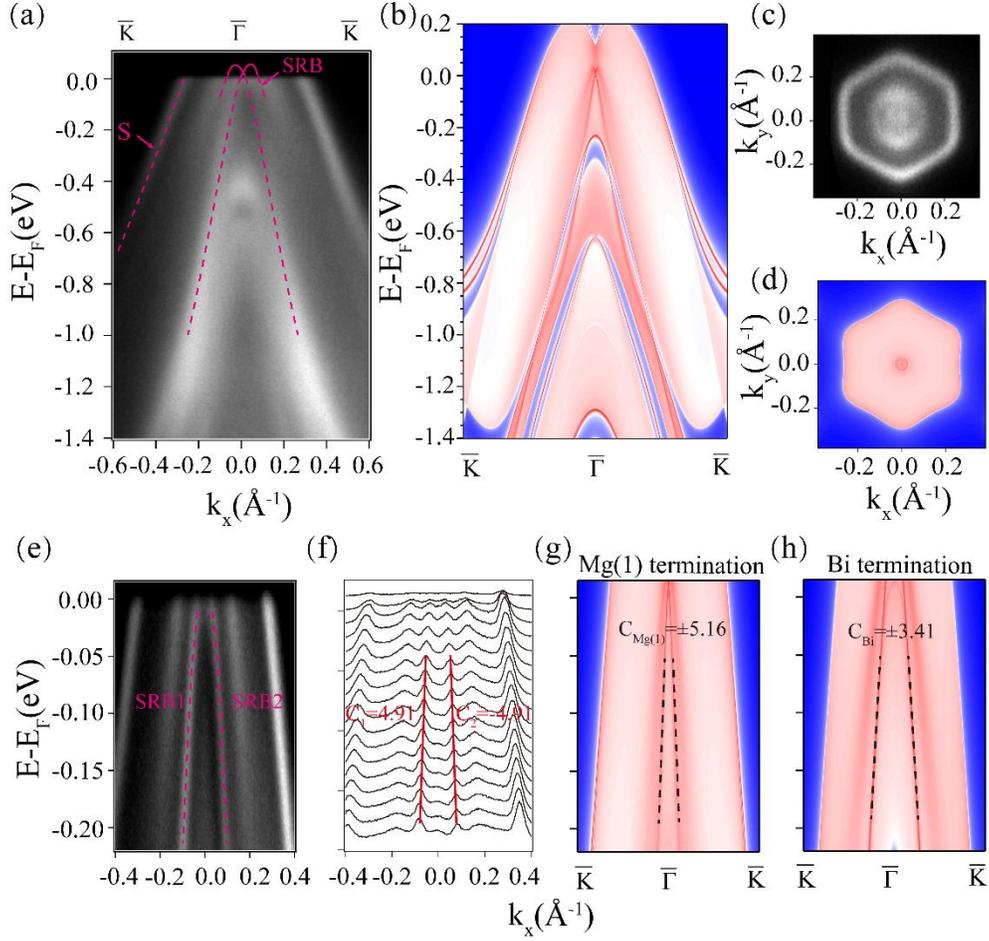

**Figure 3. ARPES measurements on Mg₃Bi₂.** (**a**) ARPES spectrum along $\overline{\Gamma}-\overline{K}$ direction and (**b**) the corresponding calculated spectrum with Mg(1)-termination. (**c**) and (**d**) Fermi surface contour from ARPES and calculation. (**e**) ARPES spectrum near Fermi level in which the SRB and S band are more pronounced, the SRB1 and SRB2 are marked by red slash line. (**f**) The corresponding MDCs to (e). The SRB slope between -0.08~-0.22eV of the energy is obtained by fitting their peak position. The slope of SRB1 and SRB2 are $C_1=4.91$ and $C_2=-4.91$, respectively. (**g**) and (**h**) the corresponding calculated spectrum of Mg(1) termination and Bi termination. The slope of SRB1 and SRB2 are ±5.16 and ±3.41 for Mg(1)- and Bi-terminated surface.

To directly confirm the TSSs of $Mg_3Bi_2$ films, we performed *in-situ* ARPES measurements at 12 K. Figure 3a shows the experimental band structure in the $\overline{K}-\overline{\Gamma}-\overline{K}$ direction. A Rashba-type splitting structure could be clearly identified, as shown by the guidelines in Figure 3a, indicating an Mg-termination sample surface. What is more, the sharply dispersive bands marked as S matches well with the theoretical calculation for Mg(1)-termination, as shown in Figure 3(b). The constant energy contour (CEC) at Fermi energy for the experimental and theoretical calculation are compared in Figure 3c and d, in which they match quite well with each other, despite the fact that the bulk continuum of the valence bands dominates the spectra weight in Figure 3d. Therefore, we have preliminarily confirmed that our $Mg_3Bi_2$ films are with the Mg(1)-termination which is different from the reported Bi-termination for single crystal $Mg_3Bi_2$.[15] ARPES spectrum near Fermi level is shown in Figure 3e, in which the SRB and S band are more pronounced. Figure 3f depicts its corresponding Momentum Distribution Curves (MDCs). A Lorentzian-type peak fit procedure was applied to each curve between -0.08~-0.22eV of the energy to find the peak position of SRB1 and SRB2. Their peaks position was forward linear fitted and thus we obtained the slope of SRB1 and SRB2 which are $C_1$=4.91 and $C_2$=-4.91, respectively. The detailed fitting result is shown in Figure S1. Figure 3g and h show the corresponding calculated spectrum of Mg(1) termination and Bi termination. The slope of SRB1 and SRB2 are ±5.16 and ±3.41 for Mg(1)- and Bi-terminated surface. Obviously our measured SRB slope is consistence with Mg(1) termination. Thus, by comparing the slope of SRB from ARPES data with calculated results, we further confirm that our $Mg_3Bi_2$ film is Mg(1)-terminated.

**Weak anti-localization effects**

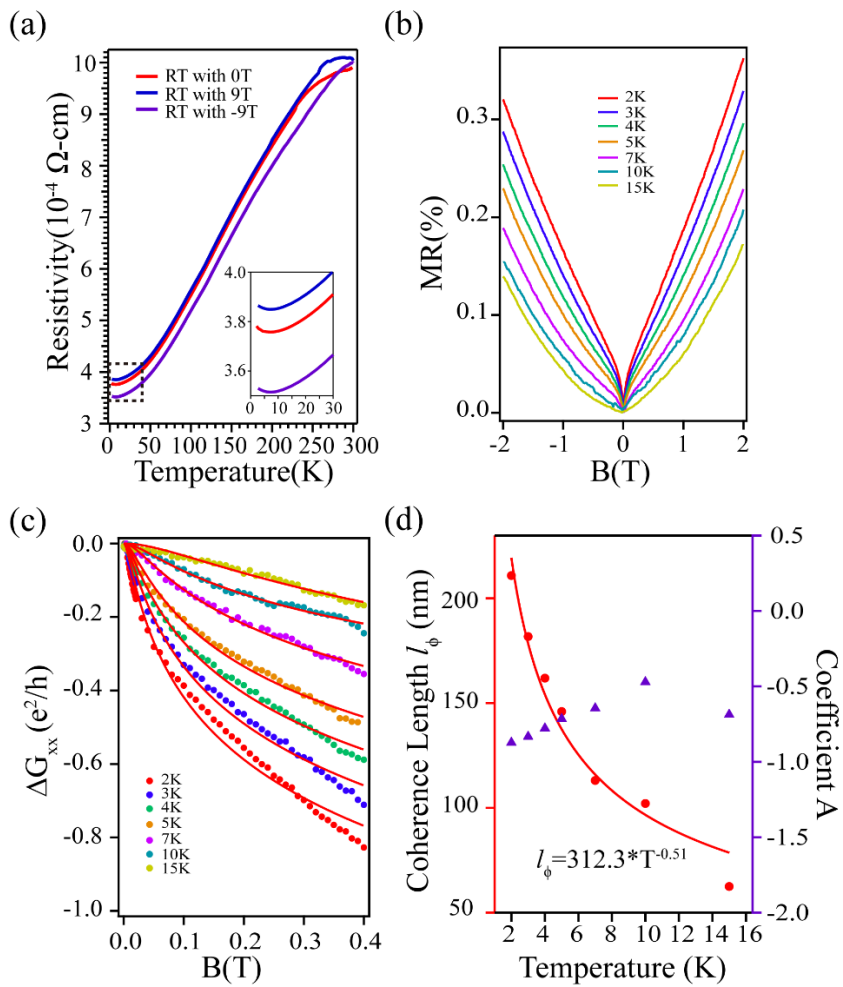

**Figure 4. Weak anti-localization effect in Mg₃Bi₂ film. (a)** Temperature-dependent resistivity of Mg₃Bi₂ at zero and ±9 T magnetic field. Inset: detail with enlarged scale of the black box. **(b)** Normalized magneto-resistance at different temperatures. **(c)** Magneto-conductance in perpendicular magnetic field configuration measured at various temperatures and their best fits to the HLN equation. **(d)** The extracted coherence length $l_\phi$ (left axis) and coefficient A (right axis) from the HLN fitting under different temperatures. The $l_\phi$ fits well with $l_\phi \propto T^{-0.51}$.

Magneto-transport is another way to study the TSSs. For transport measurements, we grew

Mg₃Bi₂ films on Al₂O₃(0001) substrate to eliminate the substrate effect. The characterization of $Mg_3Bi_2$ films on $Al_2O_3$ is shown in Figure S2. The temperature dependent longitudinal resistivity $\rho_{xx}$ of the $Mg_3Bi_2$ film is shown in Figure 4a, in which the applied magnetic field (±9 T) is perpendicular to the electrical field. $\rho_{xx}$ keeps decreasing from room temperature to 8 K, which is consistent with the behavior of semimetals. As the temperature drops below 8 K, $\rho_{xx}$ experiences an upturn (inset). This phenomenon was also observed in other topological materials and was explained as the freezing out of bulk carriers which may be caused by the presence of Coulomb interaction or *e-e* correlation among surface states on the surface of $Mg_3Bi_2$ due to disorder.[25] Figure 4b demonstrates the temperature dependent evolution of the normalized magneto-resistance (MR) which is defined as $\mathtt{MR} = \frac{R(B) - R(0)}{R(0)} \times 100\%$, where R(B) and R(0) are the resistances measured under magnetic field of B and zero field, respectively. A characteristic dip that hallmarked the existence of week anti-localization (WAL) exists at low temperature. Theoretically, the WAL phenomena for perpendicular magnetic field could be explained by Hikami-Larkin-Nagaoka (HLN) model.[26] To confirm the magneto-transport behaviors under perpendicular field, we adopt the HLN model to fit the transport data. The HLN model can be described as

$$\Delta G_{xx}(B) \equiv G(B) - G(0) \cong A \frac{e^2}{2\pi^2 \hbar} \left[ \Psi(\frac{1}{2} + \frac{B_\phi}{B}) - \ln(\frac{B_\phi}{B}) \right]$$

where G(B) is the magneto-conductance (i.e., inverse of R(B)), A is the coefficient of WAL, *e* is the electronic charge, $\hbar$ is the reduced Planck constant, $\Psi$ is the digamma function, $B_\phi = \hbar/(4el_\phi)$ is a characteristic magnetic field and $l_\phi$ is the phase coherence length. Figure 4c depicts the evolution of $\Delta G_{xx}$ versus temperature, together with the fitting results by the HLN model. The extracted coherence length $l_\phi$ as a function of T is shown in Figure 4d. $l_\phi$ decreases from 210.9 nm at 2 K to 62.5 nm at 15 K, following a power law dependence on T as $l_\phi \propto T^{-0.51}$. This monotonic decrease behavior was also observed in other topological aterials,[27] where $l_\phi \propto T^{-0.5}$ and $l_\phi \propto T^{-0.75}$ for 2D and 3D systems, respectively. The power factor of -0.51 for our $Mg_3Bi_2$ system indicates that the WAL mainly originated from the 2D topological surface states at low temperature. The WAL coefficient A at different temperatures is also shown in Figure 4d (right axis), where A=N$\alpha$ and $\alpha$ equals to -0.5 for the TSSs of symplectic universality class and N is the number of independent conducting channels. [26,28] Multiple channels could be involved in the transport due to the existence of different surface states, as well as possible conducting channels from bulk states.[29-31] The value A is between -0.87 and -0.47 for different temperatures which means N=1 or 2 indicating one or two TSS contributes as the conducting channel.

**Conclusion**

In summary, we report the growth of high quality $Mg_3Bi_2$ crystalline films on Graphene/6H-SiC(0001) substrate. The existence of topological surface states in $Mg_3Bi_2$ was confirmed by first principles calculations, ARPES and magneto-transport measurements. The magneto-transport measurements exhibit 2D WAL effect in the low magnetic field regime, which fits well with the HLN model. The coherence lengths show power law dependent on T in a relationship of $l_\phi \propto T^{-0.51}$, suggesting the main contribution from TSSs which provides direct experimental evidence for the TSSs. It was predicted that by replacing Bi by Sb or As would tune the SOC strength of $Mg_3Bi_2$ and result in a reduce spin-orbit band gap in $Mg_3Bi_{2-x}Sb(As)_x$ alloys,[15] which offers the possibilities of studying the exotic topological nature of type-II nodal line fermions. MBE techniques that we have adopted in our current work would offer a solution to the above mentioned idea of realizing a possible type-II NLS.

ASSOCIATED CONTENT

**Supporting Information**.

The methods on sample preparation, measurements detail, the detailed fitting procedure on SRB slope and characterization of $Mg_3Bi_2$ on $Al_2O_3$. Supporting information is available free of charge.

AUTHOR INFORMATION

**Corresponding Author**

*E-mail: oscarwang2008@sina.com; tjiang@nudt.edu.cn; laixinchun@caep.cn.

**Author Contributions**

The manuscript was written through contributions of all authors. All authors have given approval to the final version of the manuscript. †These authors contributed equally to this paper.

**Notes**

The authors declare no competing financial interest.

ACKNOWLEDGMENT

We thank Quan Sheng Wu and Chang-Ming Yue for helpful discussions. This work was supported by the Science Challenge Project (NO.TZ2016004), the Opening Foundation of State Key Laboratory of High Performance Computing (201601-02), the Foundation of President of CAEP (NO.201501040), the NSF of Hunan province (Grants No.2016JJ1021), the National Basic Research Program of China (NO.2015CB921303 and NO.2012YQ13012508), the General

Program of Beijing Academy of Quantum Information Sciences(Project No.Y18G17), the Youth talent lifting project (Grant No. 17-JCJQ-QT-004).

REFERENCES

(1) Qi, X.-L.; Zhang, S.-C. Topological insulators and superconductors. *Rev. Mod. Phys.* **2011**, 83, 1057-1110.

(2) Hasan, M. Z.; Kane, C. L. Colloquium: Topological insulators. *Rev. Mod. Phys.* **2010**, 82, 3045.

(3) Ando, Y.; Fu, L. Topological Crystalline Insulators and Topological Superconductors: From Concepts to Materials. *Annu. Rev. Condens. Matt. Phys.* **2015**, 6, 361-381.

(4) Fu, L.; Kane, C. L.; Mele, E. J. Topological insulators in three dimensions. *Phys. Rev. Lett.* **2007**, 98(10), 106803.

(5) Armitage, N. P.; Mele, E. J.; Vishwanath, A. Weyl and Dirac semimetals in three-dimensional solids. *Rev. Mod. Phys.* **2018**, 90, 015001.

(6) Wen J.; Guo H.; Yan C. H.; Wang Z.-Y.; Chang K.; Deng P.; Zhang T.; Zhang Z.-D.; Ji S.-H.; Wang L.-L.; He K.; Ma X.-C.; Chen X.; Xue Q.-K.; Synthesis of semimetal $A_3Bi$ (A=Na, K) thin films by molecular beam epitaxy. *Appl. Surf. Sci.* **2015**, 327,213-217.

(7) Geim A. K.; Novoselov K. S.; The rise of graphene. *Nat. Mater*. **2007**, 6(3),183-91.

(8) Seo, J.; Roushan, P.; Beidenkopf, H.; Hor, Y. S.; Cava, R. J.; Yazdani, A.; Transmission of topological surface states through surface barriers. *Nature*, **2010**, 466(7304), 343.

 (9) Yao, G.; Luo, Z.; Pan, F.; Xu, W.; Feng, Y. P.; Wang, X.; Evolution of topological surface states in antimony ultra-thin films. *Sci. Rep.* **2013**, 3, 2010.

(10) Zhu, X. G.; Stensgaard, M.; Barreto, L.; Silva, W. S. E.; Søren Ulstrup; Michiardi, M.; Bianchi, M.; Dendzik, M.; Hofmann, P.; Three dirac points on the (110) surface of the topological insulator $Bi_{1-x}Sb_x$. *New J. Phys.* **2013**, 15, 103011.

 (11) Gopal, R. K.; Singh, S.; Chandra, R.; Mitra, C.; Weak-antilocalization and surface dominated transport in topological insulator $Bi_2Se_2Te$. *AIP Adv.* **2015**, 5, 047111.

 (12) Zheng, G.; Lu, J.; Zhu, X.; Ning, W.; Han, Y.; Zhang, H.; Zhang, J.; Xi, C.; Yang, J.; Du, H.; Yang, K.; Zhang, Y.; Tian M.; Transport evidence for the three-dimensional Dirac semimetal phase in $ZrTe_5$. *Phys. Rev. B* **2016**, 93, 115414.

(13) Li, S.; Yu, Z. M.; Liu, Y.; Guan, S.; Wang, S. S.; Zhang, X.; Yao, Y.; Yang, S. A.; Type-II nodal loops: theory and material realization. *Phys. Rev. B* **2017**, 96, 081106.

(14) Zhang X.; Jin L.; Dai X.; Liu, G.; Topological Type-II Nodal Line Semimetal and Dirac Semimetal State in Stable Kagome Compound $Mg_3Bi_2$. *J. Phys. Chem. Lett.* **2017**, 8, 4814.

 (15) Chang, T. R.; Pletikosic, I.; Kong, T.; Bian, G.; Huang, A.; Denlinger, J.; Kushwaha, S. K.; Sinkovic, B.; Jeng, H.-T.; Valla, T.; Xie, W.; Cava, R. J.; Realization of a type-II nodal-line semimetal in $Mg_3Bi_2$. *Adv. Sci.* **2018**, 0, 1800897.

(16) Tamaki H.; Sato H. K.; Kanno T.; Isotropic Conduction Network and Defect Chemistry in $Mg_{3+\delta}Sb_2$-Based Layered Zintl Compounds with High Thermoelectric Performance. *Adv. Mater.* **2016**, 28, 10182.

(17) Xin, J.; Li, G.; Auffermann, G.; Borrmann, H.; Schnelle, W.; Gooth, J.; Zhao, X.; Zhu, T.; Felser, C.; Fu, C.; Growth and transport properties of $Mg_3X_2$ (x = Sb, Bi) single crystals. *Mater. Tod. Phys*. **2018**, 7, 61-68.

(18) Chang, C.-Z.; Zhang, J.; Feng, X.; Shen, J.; Zhang, Z.; Guo, M.; Li, K.; Ou, Y.; Wei, P.; Wang, L.-L.; Ji, Z.-Q.; Feng, Y.; Ji, S.; Chen, X.; Jia, J.; Dai, X.; Fang, Z.; Zhang, S.-C.; He, K.; Wang, Y.; Lu, L.; Ma, X.-C.; Xue, Q.-K.; Experimental observation of the quantum anomalous Hall effect in a magnetic topological insulator. *Science* **2013**, 340, 167.

(19) Watson L. M.; Marshall C. A. W.; Cardoso C. P.; On the electronic structure of the semiconducting compounds $Mg_3Bi_2$ and $Mg_3Sb_2$. *J. Phys. F: Metal Phys.* **1984**, 14, 113.

(20) Marzari N.; Vanderbilt D.; Maximally localized generalized Wannier functions for composite energy bands. *Phys. Rev. B* **1997**, 56, 12847.

(21) Souza I.; Marzari N.; Vanderbilt D.; Maximally localized Wannier functions for entangled energy bands. *Phys. Rev. B* **2001**, 65, 035109.

(22) Mostofi, A. A.; Yates, J. R.; Lee, Y. S.; Souza, I.; Vanderbilt, D.; Marzari, N.; Wannier90: a tool for obtaining maximally-localised wannier functions. *Comp. Phys. Comm.* **2008**, 178, 685-699.

(23) Wu, Q. S.; Zhang, S. N.; Song, H.-F.; Troyer, M.; Soluyanov, A. A.; Wanniertools: an open-source software package for novel topological materials. *Comp. Phys. Comm.* **2018**, 224, 405.

(24)Narayan, A.; Rungger, I.; Sanvito, S.; Topological surface states scattering in antimony. *Phys. Rev. B* **2012**, 86, 20140.

(25) Gorai, P.;  Toberer, E. S.; Stevanovic, V.; Effective n-type doping of $Mg_3Sb_2$ with group-3 elements. *J. Appl. Phys.* **2019**, 125, 025105.

(26) Hikami, S.; Larkin, A. I.; Nagaoka, Y.; Spin-orbit interaction and magnetoresistance in the two dimensional random system. *Prog. Theor. Phys.* **1980**, 63, 707-710.

(27) Bao, L.;  He, L.; Meyer, N.; Kou, X.; Zhang, P.; Chen, Z.-G.; Fedorov, A. V.; Zou, J.; Riedemann, T. M.; Lograsso, T. A.; Wang, K. L.; Tuttle, G.; Xiu, F.; Weak Anti-localization and Quantum Oscillations of Surface States in Topological Insulator $Bi_2Se_2Te$. *Sci. Rep.* **2012**, 2, 726.

(28) Dybko, K.; Mazur, G. P.; Wolkanowicz, W.; Szot, M.; Dziawa, P.; Domagala, J. Z.; Wiater, M.;  Wojtowicz, T.; Grabecki, G.; Story, T.; Probing spatial extent of topological surface states by weak antilocalization experiments. arXiv:1812.08711, **2018**, 1.

(29) Liu, M.; Chang, C. Z.; Zhang, Z.; Zhang, Y.; Ruan, W.; He, K.; Wang, L.-L.; Chen, X.; Jia, J.-F.; Zhang, S.-C.; Xue, Q.-K.; Ma, X.; Wang, Y.; Topological insulators in the two-dimensional limit. *Phys. Rev. B* **2011**, 83, 165440.

(30) Spirito, D.; Di Gaspare, L.; Evangelisti, F.; Di Gaspare, A.; Giovine, E.; Notargiacomo, A.; Weak antilocalization and spin-orbit interaction in a two-dimensional electron gas. *Phys. Rev. B* **2012**, 85, 235314.

(31) Liao, J.; Ou, Y.; Feng, X.; Yang, S.; Lin, C.; Yang, W.; Wu, K.; He, K.; Ma, X.; Xue, Q.-K.; Li, Y.; Observation of anderson localization in ultrathin films of three-dimensional topological insulators. *Phys. Rev. Lett.* **2015**, 114, 216601.

Supporting Information for

# Experimental Evidence of the Topological Surface States in Mg$_3$Bi$_2$ Films Grown by Molecular Beam Epitaxy

Tong Zhou[1,2,3,4,†], Xie-Gang Zhu[2,4,†], Mingyu Tong[5,†], Yun Zhang[2,4], Xue-Bing Luo[2,4], Xiangnan Xie[1], Wei Feng[2,4], Qiuyun Chen[2,4], Shiyong Tan[2,4], Zhen-Yu Wang[1,3,7,*], Tian Jiang[1,5,*], Xin-Chun Lai[2,*], and Xuejun Yang[1,6]

[1]*State Key Laboratory of High Performance Computing, College of Computer, National University of Defense Technology, Changsha 410073, P. R. China*

[2]*Science and Technology on Surface Physics and Chemistry Laboratory, Jiangyou 621908, Sichuan, China*

[3]*National Innovation Institute of Defense Technology, Academy of Military Sciences PLA China, Beijing 100010, P. R. China*

[4]*Institute of Materials, China Academy of Engineering Physics, Mianyang 621700, Sichuan, China*

[5]*College of Advanced Interdisciplinary Studies, National University of Defense Technology, Changsha 410073, P. R. China*

[6]*Academy of Military Sciences PLA China, Beijing 100010, P. R. China*

[7]*Beijing Academy of Quantum Information Sciences, Beijing, 100084, P.R. China*

†*These authors contributed equally to this paper.*

*\*E-mail: oscarwang2008@sina.com; tjiang@nudt.edu.cn; laixinchun@caep.cn.*

METHODS

Sample preparation

The $Mg_3Bi_2$ films were grown on $2\times10$ mm 6H-SiC(0001) substrate with epitaxially grown Graphene by MBE. The base pressure was maintained at $3\times10^{-11}$ mbar and the vacuum was better than $2\times10^{-10}$ mbar during growth. Mg (3N) and Bi (5N) sources were thermally evaporated from standard Knudsen cells. Mg and Bi were kept at 380°C and 540°C and with the flux rate of 0.661Å/s and 0.112Å/s (measured by Quartz crystal micro-balance), respectively. The flux ratio between Mg and Bi should be at least ~5:1 to minimize Mg vacancies and ensure the high quality of the as-grown films. The substrate was kept at 350°C during the growth and the quality of the film was monitored by *in-situ* RHEED.

Measurements

For the ARPES measurements, the spectra are excited by the He I$\alpha$ (21.2 eV) resonance line of a commercial Helium gas discharge lamp. The light is guided to the analysis chamber by a quartz capillary. In virtue of the efficient three-stage differential pumping system, the pressure in the analysis chamber is better than $2.0\times10^{-10}$ mbar during our experiments. A VG Scienta DA30L energy analyzer is used to collect the photoelectrons. $Mg_3Bi_2$ film with the thickness of 50nm was transferred *in-situ* into the ARPES chamber and measurements were done at 12K. The magneto-transport measurements were performed with the standard four-probe technique using silver paint as contacts by Physical Property Measurement System (PPMS-9). The samples for transport measurements is about 50 nm thick in a rectangular shape (5 mm $\times$ 6 mm) grown on $Al_2O_3$(0001) substrate.

The detailed fitting procedure on SRB slope

Considering the symmetry of SRB1 and SRB2 around $\bar{\Gamma}$, we fit the band by the equation of y=a×|x|+b, where |x| represent the absolute value of x. From Figure S1 we can see that the SRB band between -0.08~-0.22eV of the energy fits well by y=-4.908×|x|+0.191, and we thus conclude that the slope of SRB band is C=±4.91.

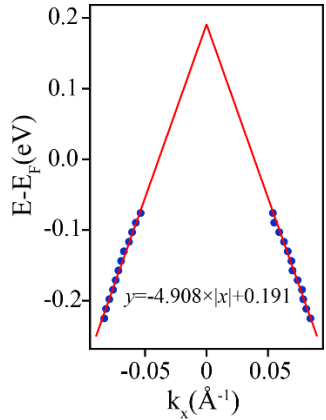

**Figure S1**. **The fitting method for SRB slope.** The blue circles indicate SRB band extracting from the MDC peaks and the red line depict the fitting results.

Characterization of $Mg_3Bi_2$ on $Al_2O_3$

For magneto-transport measurements, we grew $Mg_3Bi_2$ films on $Al_2O_3(0001)$ substrate by MBE to exclude the substrate conducting effects. The characterization of $Mg_3Bi_2$ on $Al_2O_3$ is shown in Figure S2. The in-plane lattice constant of $Al_2O_3$ is 4.758 Å while that of $Mg_3Bi_2$ is 4.677 Å, which means there is very little mismatch to grow $Mg_3Bi_2$ on $Al_2O_3$. The sharp RHEED streaks prove that and show high quality of $Mg_3Bi_2$ film (Figure S2 a and b). XRD patterns represents the (001), (002), (003), (004) and (005) peaks of $Mg_3Bi_2$, which indicates that the films grow along the *c*-axis on $Al_2O_3$ (Figure S2 c).

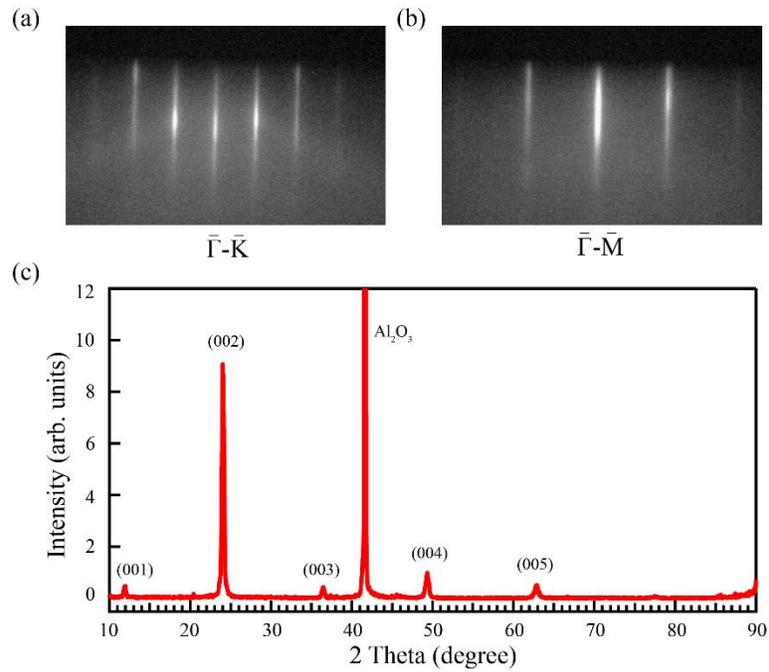

**Figure S2**. **Characterization of Mg₃Bi₂ film on Al₂O₃.** (**a**) and (**b**) The RHEED patterns of Mg₃Bi₂ film grown on Al₂O₃(0001) substrate, with the incident electron beam along the $\bar{\Gamma}-\bar{K}$ and $\bar{\Gamma}-\bar{M}$ directions, respectively. (**c**) The XRD spectra of Mg₃Bi₂ on Al₂O₃. The sharp RHEED streaks and XRD spectra are indications of high quality of Mg₃Bi₂ film on Al₂O₃.